\documentstyle[10pt,aaspp4]{article} 

\def\mincir{\raise - 
2.truept\hbox{\rlap{\hbox{$\sim$}}\raise5.truept \hbox{$<$}\ }}
\def\magcir{\raise -
2.truept\hbox{\rlap{\hbox{$\sim$}}\raise5.truept \hbox{$>$}\ }}
\def\asymp{\raise -4.3truept\hbox{$ \ \widetilde{\phantom{xy}} \ $}}
\def\MAP{{\sl MAP }}
\def\rms{{\em rms }}
\def\etal{{\em et al. }}

\begin{document}
\baselineskip=0.45truecm

\title{Adding Long Wavelength Modes to an $N$-Body Simulation}

\author{Giuseppe Tormen\altaffilmark{1}}
\affil{Cambridge University, I.o.A., Madingley Road,
Cambridge CB3 OHA, England. Email: bepi@ast.cam.ac.uk}

\and 

\author{Edmund Bertschinger}
\affil{Department of Physics, MIT, Cambridge, MA 02139 USA.
Email: bertschinger@mit.edu}

\altaffiltext{1}{Present Address: Max-Plank-Institut f\"{u}r 
Astrophysik, Karl-Schwarzschild-Strasse 1, 85740 Garching, Germany.
Email: bepi@mpa-garching.mpg.de}

\vspace{.5cm}

\centerline{{\em Accepted for publication in the Astrophysical Journal}}

\begin{abstract}
We present a new method to add long wavelength power
to an evolved $N$-body simulation, making use of the Zel'dovich
(1970) approximation to change positions and velocities of particles. 
We describe the theoretical framework of our technique
and apply it to a P$^3$M cosmological simulation performed on a cube
of $100$ Mpc on a side, obtaining a new ``simulation" of $800$ Mpc
on a side. We study the effect of the power added by long waves by mean
of several statistics of the density and velocity field, and suggest
possible applications of our method to the study of the 
large-scale structure of the universe.
\end{abstract}

\parskip=0.1truecm

\section{Introduction}
\label{intro}
Computer simulations of large scale structure play a fundamental
role in cosmology by providing a better understanding of the many 
issues related to structure formation.
The usual setup of an $N$-body simulation can be summarized as follows.
One generates some initial conditions for the simulation by
placing $N^3$ particles in a cubic box of side $L$.
The recipe for assigning an initial position and velocity to each 
particle is usually the Zel'dovich (1970) approximation.
This approximation lets one 
distribute the particles in the box so that they trace some initial 
fluctuating density field with power spectrum $P(k)$.
If the density fluctuations form a Gaussian random field, as is 
usually assumed, then $P(k)$ together with the evolution of the
expansion scale factor uniquely specifies the cosmological 
model of the simulation.
One then evolves this self gravitating system by numerically 
integrating the trajectories of all the particles under their 
mutual gravitational attraction.

Periodic boundary conditions are commonly imposed on the box. 
Since a periodic function has a discrete Fourier transform, 
the periodic boundary conditions on the box imply a discrete sampling
of $P(k)$: the only density fluctuations present in the evolution 
are those with wavelength $\lambda = 2\pi/k$ satisfying the usual periodicity 
requirement: 
$k^2 = (2\pi n_x / L)^2 + (2\pi n_y / L)^2 + (2\pi n_z / L)^2$
where $n_x$,$n_y$ and $n_z$ are positive integers.

The two parameters $L$ and $N$ determine the dynamical resolution 
of the simulation. The box size $L$ fixes the force resolution
at large scales, since fluctuations on scales $\lambda >L$ are
not included in the simulation and so are missed. It also determines
the sampling resolution of the power spectrum $P(k)$: a larger $L$
means a denser sampling of fluctuations at all scales, as
$\Delta k \propto L^{-1}$.
The particle number $N^3$ instead determines the force resolution 
at small scale since, for a given $L$, it fixes the minimum 
wavelength of the fluctuations present in the initial conditions, 
via the Nyquist relation: $k_{max} = \pi N/L$, with $k = k_x$, 
$k_y$, $k_z$. Larger $N$'s also mean more particles per object of 
interest, therefore a better resolution in mass.

The ideal configuration is of course a large box size $L$ and a 
large number of particles $N^3$. Unfortunately, even with the current 
supercomputer power, computer memory and cpu time impose severe 
limitations to the values of $N^3$ one can reasonably take. 
Given the maximum number $N^3$ of particles one can afford, 
the choice generally made is therefore to specialize the
size $L$ of the box to a specific purpose. For example, 
for a study of the general properties of a cosmological model, 
where the identity and structure of single objects is of secondary 
importance compared to the overall large scale structure and
motions produced in the model, one will take a large box, 
i.e. with size $L \approx$ few $\times 100$ Mpc, to the 
detriment of the small scale resolution in force and in mass.
On the other hand, when the interest is focussed on the internal 
structure of the class of objects under investigation (e.g. galaxies
or galaxy clusters) the box size $L$ is taken much smaller, of the
order of $\approx$ few $\times 10$ Mpc. In this case there are two
kind of disadvantages. First, the missed fluctuations on scales larger
than the box are still very important for the formation of structure.
Second, since there are only $3$ independent Fourier modes associated 
with density fluctuations of scale equal to the box size $L$,
statistical fluctuations in the density and velocity fields are 
not negligible.

The missing power on large scales will cause a sort of {\em cosmic 
bias} (in the statistical sense of the term), because the number of 
high density regions, the strength of the clustering and the amplitude
of the peculiar velocities will be systematically {\em lower} than in 
the ideal case of infinite $N$ and $L$. 
The statistical fluctuations in different realizations of the
same initial $P(k)$ on scale $\lambda \simeq L$ will instead 
introduce a {\em cosmic variance} in the simulation, since the 
measures performed on the density and velocity field will fluctuate 
around their statistical mean value.
Both effects can be dramatic if the volume $L^3$ is too small for the 
statistic one is considering, and are particularly evident if one is 
interested in the peculiar velocity field, which receives important 
contributions form linear density fluctuations of very large scale.
For example, in a Standard Cold Dark Matter universe with
a dimensionless Hubble parameter $h = 0.5$, and a linear $P(k)$ 
normalized to an \rms mass density fluctuation (in spheres of 
radius $8 h^{-1}$ Mpc) $\sigma_8=1$, the linear \rms bulk velocity 
of a cube of side $100$ Mpc is still well over $500$ km s$^{-1}$.
However, a simulation of 100 Mpc on a side will have zero bulk flow 
on the same scale, since by definition the box is at rest. The missing
power on scale $\lambda > 100$ Mpc is responsible for this cosmic
bias.
A good example of cosmic variance concerns the measure of the Hubble 
constant itself. This can assume quite different values locally,
since different patches of the universe are expected to expand
at different {\em local} rates. Turner \etal  (1992) have shown that, 
for a Cold Dark Matter universe with {\em true} Hubble constant $h=0.8$,
the local Hubble constant measured out to $30 h^{-1}$ Mpc in regions
comparable to the North Galactic Cap has an estimated value 
$h = 0.5 - 1.28$ at the 95\% confidence level.

The problems listed so far are well known in the literature 
of cosmological simulations of large-scale structure. However, 
attempts to solve them by inserting in a small scale simulation
the missing large scale power have until now been limited to
corrections applied to the mean values of the statistics (like 
$P(k)$ or the bulk velocity) (\cite{cc92}, \cite{st93}), or to
the velocity field only (\cite{st95}), but not individually to the 
velocity {\em and} density field.
In this paper we propose a new method to cure these large-scale
limitations. This method is applicable if the cosmic bias 
caused by the missing large scale power is produced
by density fluctuations which are still in the linear regime.
The idea is then to use standard linear theory and the 
Zel'dovich (1970) approximation to add to each individual particle
of an {\em evolved} simulation of size $L$ a random realization 
of the power coming from wavelengths larger than the original box 
size $L$. 
 From this idea we named our method by the acronym \MAP (Mode Adding
Procedure). The \MAP corresponds to embedding the simulated
cube in a much larger (and possibly infinite) one, therefore 
increasing the volume sampled and decreasing the cosmic bias and
variance associated with it.
Although there is virtually no upper limit to the scale 
of the added fluctuations, let us call $L_{big}$ the scale of the 
largest fluctuations one reasonably wants to introduce.

A first straightforward application of this new method is the 
construction of different realizations of a very large scale 
simulation (e.g. 3000 Mpc on a side) from just one evolved medium 
scale (e.g. 200 Mpc) simulation.
From this one could extract artificial redshift surveys of size 
comparable to the real surveys completed or in program, e.g. the 
Las Campanas Redshift Survey (\cite{lan96}), the ESO 
Slice Project (\cite{ve94}), the Sloan Digital Sky Survey (\cite{gw94}) 
and others.
The resulting simulations would have at the same time a
scale sufficient to address the issue, and enough resolution on small
scale to properly identify galactic halos.
The artificial surveys can be used to calibrate the different
sources of uncertainties present in the real data (sparse sampling,
redshift errors etc.), to estimate the scientific impact expected from
the new data, and to compare them with the predictions of different
cosmological models, via a number of statistics. 
Other possible applications include the dynamical study of the role 
played in the structure formation process by linear density
fluctuations on very large scale. Do they couple with nonlinear modes 
during time evolution? How do they trigger the formation of cosmic 
structure on all scales? How important are they when studying the 
velocity field or superclustering phenomena? 

The paper is organized as follows. Section 2 is a description of 
our method. In Section 2.1 we discuss the Fourier space manipulation 
required to apply our method to a simulation. Section 2.2 describes 
the corresponding steps performed in position space: a Mode Removing
step and a Mode Adding step.
In Section 2.3 Lagrangian and Eulerian ways of implementing 
the technique are considered and discussed. 
In Section 3 we apply the method to an $N$-body simulation
and compare some statistics of the density and velocity field
obtained from the \MAP simulation with the same statistics applied
to a real $N$-body simulation performed on large scale.
Section 4 gives a summary of the results and presents some conclusions.

\section{Method}
\label{met}

\subsection{Fourier Space Manipulation}
\label{fsm}

The Fourier space of a periodic simulation can be thought as a cubic 
lattice. The sampling of the initial $P_\delta(k)$ is made on a regular 
cubic grid centered on $\vec k =\vec 0$, ($|\vec k |\equiv 2\pi/\lambda$), 
with inter-grid size $\Delta k=k_{min}=(2\pi /L)$ and extension in 
every direction determined by the number $N^3$ of particles used, 
via the Nyquist relation: $k_{max}=\pi N/L$.

\placefigure{fig1}

The diagram in Fig. 1a shows the central region of this sampling.
The dots correspond to the positions where $P_\delta(k)$ is evaluated,
and the regular grid divides the Fourier space in cubes of equal side
$\Delta k$.  Each cube is associated with a discrete Fourier component 
$\hat\delta_{\vec k}$ of the density fluctuation 
field $\delta(\vec x)$: the intensity of density fluctuations 
of wavenumber $\vec k$ is the mean power per mode
$\langle |\hat\delta_{\vec k}|^2\rangle \approx P_\delta(k) (\Delta k)^3$.
Including in the simulation density fluctuations on scales 
$\lambda > L$ corresponds to improving the sampling of $P_\delta(k)$ 
around $\vec k=\vec 0$. Our scheme is thus a kind of mesh-refinement 
algorithm implemented in Fourier space.

Our approach is the following.
We first remove, around $\vec k = \vec 0$, the power associated 
with the Fourier modes of the original sampling. This means deleting the
power of a number of cubes each of side $\Delta k = (2\pi/L)$ in the central
region of Fourier space. We then add back new power by filling the same
region with a larger number of smaller cubes, each of side 
$\Delta k^\prime = (2\pi/mL)$ (with $m$ a positive integer). 
The power per mode of these new cubes is assigned with a random 
realization of the (Rayleigh distributed) linear power spectrum
$P_\delta(k)$ at the corresponding positions; this ensures that 
the correct amount of power is added to the simulation. 
The procedure is sketched in Fig. 1b for the case $m=4$: a grid four 
times finer is substituted for the original one in Fourier space, 
out to an extension $r_k \equiv k/k_{min}=1$ in each direction.
Subtracting a cubic region of extension $r_k$ corresponds to 
removing $(2r_k+1)^3$ cubes of side $2\pi/L$ from the Fourier space. 

Our method adds long-wavelength power to a simulation at the end of 
time evolution. To ensure that the result is dynamically consistent
we must remove and add only power associated with fluctuations that
are still in the linear regime. The linearity constraint may be
interpreted in different ways. The most general
requirement is that the root mean square density fluctuation associated 
with linear waves must be smaller than unity: $\sigma_\delta < 1$.
Another straightforward characteristic of linear waves is that they evolve 
in agreement with the equations of linear theory.
A third requirement is that linear waves should not dynamically
couple with any other wave. This issue has been explored by Jain and
Bertschinger (1994),  who find that for a cold dark matter spectrum of
density perturbations mode coupling can transfer significant power
to shorter waves from long wavelengths still nearly in the linear regime.
We can monitor this by checking that the longest mode in our original
box has linearly growing amplitude, but even this is not a rigorous
limit because longer waves absent from the box might have caused the
amplitude already to depart from linear growth.  For now we adopt the
practical viewpoint of trying the method and later testing for effects
of nonlinear coupling.  We will discuss this issue further in the
summary at the end of this paper.

Whichever linearity requirement we choose, this will 
put a limit on the region of Fourier space where we can perform the power 
substitution. In the example of Fig. 1 we manipulate the power in a cube of 
extension $r_k=1$; depending on the simulation, we may extend the substituted 
region up to higher $\vec k$.

\subsection{The \MAP idea: Mode Removing and Mode Adding}
\label{idea}

To remove and add the power as described in Section 2.1 we decided 
to use the displacement field and to perform the power manipulation 
by mean of the Zel'dovich (1970) approximation. 
In this approximation each fluid element moves along a straight line 
with a velocity linearly extrapolated from its initial velocity.

We will work in co-moving coordinates with $\vec q_i$ and $\vec x_i$ being
respectively the initial (Lagrangian) and final 
(Eulerian) position of the $i$-th particle of the original simulation, 
$i=1,2,\dots, N$. We will denote by $\vec x^\prime$ the final particle
positions after the mode removing step and $\vec x^{\prime\prime}$ the 
final positions after the mode adding step. 

Before describing how we perform the mode substitution in practice,
we will briefly review some relations between the density, 
velocity and displacement fields that will be used later.
The relation between the Eulerian and Lagrangian position of a
fluid element at $\vec x$ is given by the displacement field 
$\vec \psi$: $\vec x(\vec q,t)=\vec q + \vec \psi(\vec q,t)$.
By applying mass conservation and assuming that $\vec x(\vec q\,)$ is 
one-to-one (no orbit mixing) we can write the exact relation
\begin{equation}
\delta(\vec x) = 
\left\| {\partial\vec x \over \partial \vec q}\right\|^{-1} - 1 =
-\vec \nabla_q \cdot \vec\psi(\vec q\,) + {\cal O}(\psi^2).
\label{lin1}
\end{equation}
Note that $\vec \psi$ is the full, nonlinear displacement.
This relation is still mixing Lagrangian and Eulerian coordinates;
however, in the linear approximation we may consider the displacement 
to be a function of $\vec x$ and write
\begin{equation}
\delta^{\,(1)}(\vec x) = -\vec \nabla_x \cdot \vec\psi^{\,(1)}(\vec x)
\label{lin2}
\end{equation}
where now both $\delta^{\,(1)}(\vec x)$ and $\vec \psi^{\,(1)}(\vec x)$ 
are first order quantities.
We see how the time dependence of the linear displacement
field is that of $\delta^{\,(1)}(\vec x,t)$, that is, considering only the
growing mode $D_+(t)$: $\vec \psi^{\,(1)}(t)\propto D_+(t)$.
Using this result and taking the time derivative of the mapping 
$\vec q \longrightarrow \vec x(\vec q\,)$ we obtain to first order the 
linear relation between the co-moving peculiar velocity 
$\vec v \equiv d\vec x/dt$ and the displacement field at time $t$:
\begin{equation}
\vec v^{\,(1)}(\vec x) = H_0 f(\Omega) \vec \psi^{\,(1)}(\vec x)
\label{lin4}
\end{equation}
where $f(\Omega)=d\log D_+/d\log a\approx \Omega^{0.6}$. In particular,
if $\Omega=1$ the linear displacement field and the linear peculiar velocity 
field coincide if units of km s$^{-1}$ are used.
The Zel'dovich (1970) approximation is then written as:
\begin{equation}
\vec x(\vec q\,)= \vec q + D_+(t)\vec \psi^{\,(1)}(\vec q\,)=
\vec q + {D_+(t) \over H_0 f(\Omega)} \vec v^{\,(1)}(\vec q\,)\ .
\label{zel}
\end{equation}

The steps we follow in practice are the following (all equations are
meant at a given time $t$):

{\bf 1.} Starting with a simulation performed with $N$ particles on
a cubic volume $L^3$ we compute the co-moving displacements as 
$\vec \psi_i= \vec x_i - \vec q_i$, $\ \ \ i=1,\dots,N$. 
We use these displacements to define a displacement field $\vec \psi$ 
on a regular grid in position space. This can be done in different ways, 
as we will see in the next subsection.

{\bf 2. Mode Removing}: We decompose the displacement field 
into the contributions $\vec \psi^{long}$, due to the modes that we 
are going to subtract from the simulation, and $\vec \psi^{short}$ 
due to all the other modes: $\vec \psi=\vec \psi^{long} +\vec \psi^{short}$.
We subtract the long wavelength power from the simulation by 
interpolating $\vec \psi^{long}$ to the {\em Eulerian} position of each
particle and subtracting it, changing in this way each particle's position.
The velocities are changed in the same way, by taking advantage 
of the linear relation between them and the displacements Eq.(\ref{lin4}):
\begin{equation}
\vec x^\prime_i=\vec x_i - \vec \psi^{long}(\vec x_i); \ \ \ 
\vec v(\vec x^\prime_i)=\vec v(\vec x_i)- H_0 f(\Omega) 
\vec \psi^{long}(\vec x_i); \ \ \
i=1,2,\dots, N.
\label{erem}
\end{equation}

Another way to subtract the large scale contribution to the velocities
would be to directly decompose the velocity field as we did for the
displacement, instead of using the linear relation between the two. 
We expect the two procedures to give the same result, as long as
the basic assumption of removing only linear modes holds.
Note that we did not subtract the displacements $\vec\psi^{long}$ at the
$\vec q$ (Lagrangian) positions of the particles. If we did that,
we would disrupt the nonlinear structures which formed during the
evolution of the simulation, and smear out the power on small scales.
So far the change in positions does not make use of the Zel'dovich 
approximation,
since the  subtracted displacement is the actual one and not the initial one
times the growth factor $D_+$.
We are, however,
assuming linear theory in the  relation between the velocity and displacement
fields, which is only approximate even for long waves.

{\bf 3. Mode Adding}: We generate a new set of initial conditions
consisting of random long-wavelength displacements, 
$\vec \psi_1^+, \dots, \vec \psi_N^+$ in a cube of side 
$mL$, by randomly sampling the power spectrum $P_\delta(k)$ only at
those positions around $\vec k=\vec 0$ corresponding to the long
wavelength modes we are going to add.
These displacements are then linearly evolved up to the present time, 
again interpolated to the {\em Eulerian} position of each particle, 
and added to the positions and velocities as prescribed by the
Zel'dovich approximation:
\begin{equation}
\vec x^{\prime\prime}_i=\vec x^\prime_i + D_+(t)\vec \psi^+(\vec x^\prime_i); \ \ \
\vec v(\vec x^{\prime\prime}_i)=\vec v(\vec x^\prime_i)+ D_+(t) H_0 f(\Omega) 
\vec \psi^+(\vec x^\prime_i); \ \ \
i=1,2,\dots, N
\label{eadd}
\end{equation}
The final result is a set of particle positions and velocities which
now include the effect of density fluctuation waves as long as $mL$.

We stress the importance of adding the long wave power by
interpolating $\vec\psi^+$ to the new positions $\vec x'$, not at the
original positions $\vec x$. In fact, in the \MAP view the $\vec x$ 
are just some {\em wrong} Eulerian positions where the particles
stand due to the missing large scale power. Considering only
the waves we subtract and add, we can identify $\vec x^\prime_i$ with
the Lagrangian position $\vec q$. The correct way to apply
the Zel'dovich (1970) approximation is 
$\vec x=\vec q + D_+\vec \psi^+(\vec q)$, which indeed corresponds
to Equation (\ref{eadd}).
As a check we also tried the alternative formulation:
$\vec x^{\prime\prime}_i=\vec x^\prime_i + D_+(t)\vec \psi^+(\vec x_i)$.
As expected, the results do not
satisfactorily reproduce the linear long wave power 
one is introducing.

\subsection{Lagrangian vs Eulerian approach}
\label{laeu}

As sketched in Section 2.2, we need to define a displacement field
on a regular grid. This can be performed in two ways.
Starting from a set of $N$ displacements $\vec \psi_1,\dots,\vec \psi_N$
we can assign each displacement $\vec \psi_i$ to its initial position 
$\vec q_i$:
$\vec \psi_i \equiv \vec \psi(\vec q_i)$ and write:
\begin{equation}
\vec \psi_L(\vec q\,)={\sum_{i=1}^N \vec \psi_i \delta_D(\vec q-\vec q_i)
\over \sum_{i=1}^N \delta_D(\vec q-\vec q_i)}.
\label{ldisp}
\end{equation}
where $\delta_D$ is a Dirac delta function. The subscript $L$ stands for
{\it Lagrangian} because the resulting displacement field is defined on 
a regular grid of initial positions.

Alternatively, we can assign every displacement $\vec \psi_i$ to the 
corresponding final position $\vec x_i$: 
$\vec \psi_i \equiv \vec \psi(\vec x_i)$ and interpolate 
such displacements onto a regular grid of final 
positions $\vec x_g$ by mean of a suitable window function $W$; 
the resulting displacement field will be called {\it Eulerian}
and indicated by a subscript $E$:
\begin{equation}
\vec \psi_E(\vec x_g)= {\sum_{i=1}^N \vec \psi_i
W(\vec x_g, \vec x_i) \over \sum_{i=1}^N W(\vec x_g, \vec x_i)}.
\label{edisp}
\end{equation}

In subtracting the longest waves
from the parent simulation we used the Eulerian displacement field 
$\vec\psi_E(\vec x)$, with a window function $W$ corresponding to a
Triangular Shaped Cloud (TSC) interpolation on a regular cubic grid
with $32^3$ mesh points.
The mode adding part was performed instead using the Lagrangian
field $\vec\psi_L(\vec x)$, since we assigned the displacements generated by
the long waves to a grid of initial positions $\vec q$ as is usually
done when generating standard initial conditions for a simulation.
However, we do not expect the choice between $\vec\psi_L$ and 
$\vec\psi_E$ to be fundamental to the final result.
In fact, the displacement fields we are considering are due only 
to long, linear waves; 
$(\vec \psi_L -\vec \psi_E)$ at a given position is a second-order 
quantity.

In practice, the displacements employed by Equations (\ref{erem}) 
and (\ref{eadd}) are computed by interpolating the displacement 
field from the grid points to the Eulerian particle position:
\begin{equation}
\vec \psi(\vec x)= {\sum_{\vec x_g} \vec \psi(\vec x_g)
W(\vec x, \vec x_g) \over \sum_{\vec x_g} W(\vec x, \vec x_g)}
\label{edis2}
\end{equation}
where $\vec \psi$ may be either $\vec \psi_L$ or $\vec \psi_E$ and
$\vec x$ is the particle position. The sum is extended over all the 
grid points $\vec x_g$ in the simulation. If $\vec \psi_L$ is used,
the grid points $\vec x_g$ stand of course for the initial positions
$\vec q_i$, $i=1, \dots, N$.

\section{Application to a simulation: MAP8x144}
\label{sim}

As a first application of our technique we will take a medium range,
high resolution $N$-body simulation, originally evolved in a periodic
cube of side $L$, replicate it $m^3$ times in a larger cube and add
to it the missing power from the long wavelengths not sampled in the
original cube, up to $\lambda=mL$.
We use a P$^3$M $N$-body cosmological simulation of 
an Einstein-De Sitter cold dark matter universe, with 
a dimensionless Hubble constant $h=0.5$, evolved in a cube of side 
$L=100$ Mpc (\cite{gb94}). 
The simulation was run with $144^3$ collisionless particles, each with mass 
$2.3 \times 10^{10} M_\odot$ and a Plummer softening radius of $65$ kpc. 
We chose the output of the simulation corresponding to a linear 
normalization $\sigma_8=0.7$. We will refer to this simulation as P3M144.
This simulation has high mass and force resolution, and is
particularly suited for studying the dynamics of cold dark matter 
halo formation and the small to medium scale structure. 
On the other hand, its size is too small to allow a study 
of the velocity field on large scales through statistics like 
the bulk flow or the velocity correlation tensor. 
We will compare statistics of the density and velocity field 
for P3M144, both before and after applying the \MAP, and for 
a reference $N$-body simulation evolved on a much larger scale.
The latter, which we call P3M256, was run with identical cosmological 
parameters as P3M144, but has $256^3$ particles in a cube of side 
$L=640$ Mpc and a Plummer softening radius of $160$ kpc.

One way to verify the assumption of large-scale linearity for the
fields in P3M144 is to measure the \rms power associated with the 
long waves that are removed from and added back to it. The \rms 
density fluctuation and displacement are defined directly in 
Fourier space respectively as 
\begin{equation}
\sigma_\delta=\sqrt{\sum P_\delta(k) (\Delta k)^3} \ \ \ \ \hbox{and}
\ \ \ \ \psi_{rms}=\sqrt{\sum P_\delta(k) {(\Delta k)^3 \over k^2}}
\label{pow2}
\end{equation}
where the sums are extended over the modes under investigation.
Referring to the picture of Fourier space in Fig. 1, we tried the
mode substitution on P3M144 in a cubic region of extension $r_k = 1$. 
This corresponds to removing the power associated with the $27$ central 
cubes of side $\Delta k= (2\pi/100)$ Mpc$^{-1}$ each.
The displacement field associated with the removed region of Fourier 
space is what we called $\vec \psi^{long}$ in Equation (\ref{erem}).
Its root mean square value computed from P3M144 as in Eq.(\ref{pow2})
results $\psi^{long}_{rms}= 7.1$ Mpc. The \rms displacement due to all
the wavelengths present in the simulation is $\psi_{rms}=11.2$ Mpc, 
showing how most of the displacement is due to the long waves present
in the simulation. This is in agreement with what we said earlier about 
the peculiar velocities: both the velocity and displacement fields
receive the biggest contribution from large-scale density fluctuations.
This fact however does not invalidate our linear theory approximation,
because the long wavelength displacement and velocities correspond to
nearly uniform (bulk) motions for the particles, with no creation of nonlinear
structures such as pancakes. This is also confirmed by the value of 
the \rms density fluctuation due to the removed modes: 
$\sigma^{long}_\delta=0.55$ is less than one, as we would expect if 
the linear approximation applies.  We will see if this is small enough
when we compare with a larger simulation below.

We computed $P_\delta(k)$ from the simulation at different time-steps
and plot in Fig. 2a the growth rates of the lowest modes, normalized 
to the value of $P_\delta(k)$ at $a(t)=0.1$.
Fig. 2b shows the analogous plot for P3M256, for comparison. Note that
in this case the plot is normalized to a scale factor $a(t)=0.2$, and 
the most evolved output of the simulation corresponds to $a(t)=0.7$.
 From Fig. 2 we deduce that all the modes of P3M144 up to $r_k\sqrt{3}$
are approximately in the linear regime, since their growth rate departs 
from the linear prediction by less than 20\% even at the latest times.
On the basis of these two tests we conclude that $r_k=1$ is a good choice
for the region of Fourier space of P3M144 where we will carry out our mode 
substitution.

\placefigure{fig2}

We recall that the new sampling of the power spectrum around 
$\vec k=\vec 0$ will have a resolution $m$ times the initial one: 
$\Delta k^\prime=\Delta k/m$, as shown in Section \ref{fsm}.
We estimated that $m=8$, twice the resolution of the example in Fig. 1b,
is a reasonable choice.
In terms of wavelength, this roughly corresponds to
removing linear displacements generated by density fluctuations
of scale $\lambda=100$ Mpc and to adding back displacements
associated with scales $\lambda \in [67,800]$ Mpc. 
However, since the geometry of Fourier space forced us to remove and 
add power in cubic regions, there are some shorter wavelengths
removed and added as well, corresponding to the edges of the cube. 
Specifically, we remove modes up to a minimum $\lambda_{min}=58$ Mpc 
and add modes up to a minimum $\lambda_{min}=38$ Mpc.
Considering a maximum fluctuation scale of $800$ Mpc guarantees 
that we are including most of the power driving the velocity field for 
a standard cold dark matter model. In fact, a cube of side $800$ Mpc 
has a \rms bulk flow of roughly $150$ km s$^{-1}$ for $\sigma_8=1$, 
much lower than the bulk flow of the original box. 
Taking $8^3$ replicas of the original simulation blows up the total 
number of particles to more than $1.5 \times 10^9$, too many for us
to retain. We chose to keep a different random subsample of 
$32,000$ particles from each of the 512 replicas of the original simulation, 
for a total of $16,384,000$  particles over the $(800$ Mpc)$^3$ volume.
We will refer to this simulation as MAP8x144.

The \rms displacement and density fluctuation due to the added 
modes and computed as in Eq. (\ref{pow2}) are $\psi_{rms}^{+}=10.1$ 
Mpc and $\sigma_\delta^+=0.63$ (we recall that the linear normalization 
for P3M144 is $\sigma_8=0.7$).
We find as expected that the added power is larger than the 
power we subtracted from the simulation: $\psi_{rms}^{+}> 
\psi_{rms}^{long}$ and $\sigma_\delta^+ > \sigma_\delta^{long}$. 
However, the extra power has been added in a way consistent with 
the power spectrum underlying the simulation and satisfying the
requirements of the linear approximation.

The figures quoted for the \rms density fluctuations associated with
the removed and added modes might look high when compared to the
normalization of the original simulation. In particular, one might
wonder how can the long waves introduce an \rms density contrast as
big as $\sigma_\delta^+=0.63$, if $\sigma_8=0.7$.
The answer to this question is found in the different ways which
were used to compute the density contrast. While $\sigma_8=0.7$
refers to a spherical top-hat filter, the figures associated
with the long waves were computed with Eq.(\ref{pow2}), 
which gives direct summations of the power of each mode with
no filter function to smooth it. Equivalently, the filter function
for the spherical top-hat is sufficiently different from the filter 
function corresponding to the discrete mode distribution we are 
dealing with that a direct comparison of the two is not possible.

Throughout this section we will compare the results and statistics
from MAP8x144 with those from P3M144 and P3M256.
We divide the analysis in two parts. The first part is dedicated to 
the study of the density field: we will show the effect of adding 
long wavelengths on both the morphology and the statistics.
In the second part we will study the velocity field.

\subsection{The density field}

To give a first visual impression of the effect of long waves on
a distribution of matter, we applied the \MAP to a two-dimensional sheet
of particles regularly spaced. The result is shown in Figure 3: the
modulation produced by the long waves is evident. The displacement field
has moved the particles from their grid positions, creating density 
fluctuations on the required scales; the shortest fluctuations
are of the order of 40 Mpc. In the most crowded regions the particle
trajectories are relatively close to shell crossing, but have not yet 
reached it, as can be clearly seen in the right panel, which shows a 
blow up of the part of slice enclosed by the square imprinted on top of 
the left panel. The magnified region, $200$ Mpc
on a side, is subdivided into four equal parts, each of which has the same
size as P3M144. One can see how the effect of long waves on different 
replicas of the original simulation changes the global distribution of matter,
modulating the pre-existing structure in different ways on different copies.

\placefigure{fig3}

Next we want to make sure that the small scale, nonlinear clustering present
in P3M144 is not changed or disrupted when we add the extra long wavelength
power. We tested this expectation by applying the \MAP to a two dimensional
network of filaments, shown in Figure 4.
\placefigure{fig4}
After the action of the \MAP the network is still connected, i.e. the 
topology of the structure has not been modified by long-wave fluctuations.
However, the filaments have been stretched here and compressed there, along 
all three dimensions, with different strength and effect in different places.
We can measure how much ``stretching'' or ``compression'' the long waves
produced on the structure at different positions by looking at the derivative
of the displacement field along the filaments. We randomly choose one of the
16 filaments shown in Fig. 4 and computed the three components of the
displacement field along it. These are shown in the first panel of Fig. 5.

\placefigure{fig5}

The displacements plotted in Figure 5 are continuous functions
as expected, meaning that there is no ``stripping'' of two nearby particles
due to the long waves.
The total three dimensional displacement of a particle is shown in the
second panel. Its derivative tells us how much two initially 
close particles can be taken apart by the long waves. The steepest part of
the function is blown up in the third panel. The derivative on the slope
between $x=500$~Mpc and $x=520$~Mpc is approximately constant and has a value
$ds/dx_0 \approx 0.5$. If we take this as a figure representative of the
whole displacement field, it tells us that particles originally in a structure
of size $R$ will be taken apart at most by an amount equal to $R/2$, that
is half the size of the structure. Hence the identity of e.g. dark matter
halos is preserved by the \MAP if the mean inter-halo separation is bigger than
the size of each individual halo. As this is usually the case, we do not 
expect the small structures of the halo to be appreciably
disturbed by the action of
the long linear waves of the \MAP, although the final word on this issue
can only be given by direct analysis of the halos through some group
finding algorithm, which we have not yet done.

In Figure 6 we compare MAP8x144 and P3M256 by showing slices of 
$640 \times 640 \times 10$ Mpc$^3$. The linear power spectrum normalization 
is $\sigma_8=0.7$ for all slices.
In this Figure we sampled the particles of the three simulations so that 
the number of particles shown in each slice is approximately equal.
Fig. 6a shows a slice from the original P3M144, replicated over such 
a volume, prior to any mode substitution; the cut shows regions with
structure as big as the whole original simulation P3M144, and the 
periodicity over $100$ Mpc is evident. Fig. 6b shows the same Eulerian
slice after we performed the mode substitution. 
\placefigure{fig6}
In Fig. 6b most of the periodicity has been 
disrupted: the long waves have stretched here and compressed there 
the original pattern of clustering and voids, resulting in a more 
varied structure. New patterns have developed with a characteristic
scale much larger than the original box size.
Fig. 6c shows a slice from the P3M256 $N$-body simulation. 
It is relatively easy for the eye to see the richer range in patterns 
of this true simulation when compared to Fig. 6b. Such a comparison
brings to evidence the residual periodicity of MAP8x144, in the form
of cell-like structure of about the size of the original simulation,
P3M144. Cells of similar size also appear in Fig. 6c, but they are 
less evident due to their more irregular distribution.
Although the figures give an idea of the performance of the \MAP on large
scales, they are also slightly misleading in that they do not show how
much better MAP8x144 actually is on small scales owing to its better
resolution when compared to P3M256. This can only be shown by specific 
tests, like the ones we are going to present in the next Sections.
On the whole, the largest structures which can be identified in Fig. 6b 
and Fig. 6c have roughly the same size, of the order of about $150$ Mpc.
The relatively emptier regions seen in Fig. 6b, in comparison with
Fig. 6c, are due to the sparse sampling we had to apply to P3M144
(roughly one particle in 90) in order to reduce the total number of 
particles in the (800 Mpc)$^3$ box to 16 million.

\placefigure{fig7}

In Fig. 7 we plot the logarithm of the one point density fluctuation 
distribution function 
$f(\delta)$ obtained by computing the density field on a regular 
lattice with $5$ Mpc spacing using the TSC
interpolation scheme.  We also computed the \rms density fluctuation
for our various simulations; the values obtained are listed on the first
row of Table 1.
We see that MAP8x144 has a larger number of grid points 
with no particles and a longer tail of high over-densities.
The abundance of empty grid points is due to the sampling problem noticed 
already in Fig. 6b. In order to produce MAP8x144 we took only $32,000$ 
particles from each copy of P3M144. This corresponds to an average of 
$32$ particles contributing to the density value of each grid point;
such number is evidently too low to allow a good sampling of very 
under-dense regions, which turn out completely empty. 
This does not happen for P3M144, where the average number of particles 
contributing to a grid point is about $3000$. As for P3M256, 
the Figure shows that $64$ particles per grid point seem enough for a good 
sampling, but this is probably due to the larger particle mass of this
simulation.

\placetable{tab}

\begin{table}
\caption{}
{Density and Velocity Moments}
\label{tab}
\begin{center}
\begin{tabular}{||c|c|c|c||} \hline
  & P3M144 & MAP8x144 & P3M256 \\ \hline
$\langle\delta^2\rangle^{1/2}$ & 1.96& 1.94 & 2.00 \\ \hline
$\langle|\vec v|\rangle$ & 655 km s$^{-1}$& 722 km s$^{-1}$& 699 km s$^{-1}$\\ \hline
$\langle v^2\rangle^{1/2}$ & 768 km s$^{-1}$& 827 km s$^{-1}$& 804 km s$^{-1}$\\ \hline
\end{tabular}
\end{center}
\end{table}

\placefigure{fig8}

Fig. 8 shows $P_\delta(k)$ for P3M144, both before and after removing
the long waves, as well as $P_\delta(k)$ for MAP8x144 and P3M256.
The spectra have been computed using a $512^3$ regular grid ($128^3$ for
P3M144 after the mode removing); they include both a deconvolution from 
the interpolating scheme and a shot noise subtraction. 
The different starting and ending point for the curves are a consequence 
of the different scales of the simulations.

If our assumption of linearity is correct we should see in
$P_\delta(k)$ the same change in power that we performed on 
the displacement field.
In fact we do see that the amplitude of the power spectrum 
after mode removing has decreased by a factor of about $40$ in the 
first bin, corresponding to modes with $\Delta k/k_{min} \in [0.5,1.5]$.
We note however that the power spectrum after mode removing seems 
to be slightly lower than the original one: defining
$b^2(k)\equiv P_\delta(k)/P_{\delta}^-(k)$, where $P_{\delta}^-(k)$
is the power spectrum of P3M144 after the mode removing step,
we found that $b(k) \approx 1.15$ for high $k$. 
This may be due to some mode coupling between small and large $k$,
caused by the fact that some of the subtracted waves are not evolving 
in a sufficiently linear way. If this is the case, then our tolerance
of $\sim 20\%$ departures from linearity shown in Figure 2 would not 
be enough to ensure accuracy to better than 15\%.
To test this possibility one could apply the mode removing to a 
larger simulation, e.g. twice the linear  size of P3M144, and see 
if the effect is still there. 
On the other hand, the fact that P3M256 also has a power spectrum 
amplitude which is little lower than that of P3M144 on small scales
suggests that the explanation of these differences could be more
complicated.

As a check of our method we also tried to subtract from P3M144 a larger 
number of modes, corresponding to cubes of extension $r_k=2$, $3$ and $4$ 
in Fourier space. We found that removing shorter and shorter waves from 
the displacement field does not correspond to removing the equivalent 
power from the density fluctuations because the linear 
relation between the displacement and density 
fluctuation field breaks down for small scales.

The spikes shown by the power spectrum of the \MAP simulation at 
$\log\ k \in [-1,-0.6]$ are an artifact due to the uneven sampling
of power (as shown in Fig.1b). This was not taken into account in
the way $P_\delta(k)$ is numerically evaluated, since the power
summation is made in spherical shells with constant width 
$\Delta k= (2\pi/800)$ Mpc$^{-1}$. The effect shows up at values 
of $k$ where the old and new power sampling mix together, and 
disappears at higher wavenumbers due to the higher number of modes
present in each shell.

The global agreement of the power spectra of P3M144, MAP8x144 and 
P3M256 should imply an equally good agreement of the corresponding
mass autocorrelation functions. We indeed found that this statistic
differs by less than 35\% (or 0.13 in logarithm) between the three
simulations over range of pair separations not influenced by 
small scale force softening or by border effects.

\subsection{The velocity field}

We would like to test the \MAP performance on the velocity field,
in the same way as we tested the density field. Unlike the density,
the velocity has the advantage of being defined (for single particles)
without any smoothing, enabling us to study directly the distribution 
function of the raw particle velocities.
In Fig. 9 we plot the distribution function $f(v)$ of the velocity
modulus for the three simulations. The gain in peculiar velocities due 
to the power associated with long waves is evident. Table 1 lists the
first and second moment of the distributions for easy comparison. 
Besides increasing the peculiar velocities, long waves add coherence 
to the velocity field, so that the average velocity of a region of 
size 100 Mpc is roughly zero for the P3M144 simulation, but is of 
a few hundred km s$^{-1}$ for the \MAP simulation. 

\placefigure{fig9}

We would like to measure the velocity power spectrum $P_v(k)$ for the
simulations. Unfortunately, this is not a very well defined quantity. 
In fact, in order to define a velocity field on a regular grid one needs 
to take the ratio between the momentum and the density fields. 
If there are no particles in the neighborhood of a grid point, the density 
and the momentum will be zero there, and the velocity will be undetermined. 
Therefore, in order to define a velocity field at all grid points 
one has to smooth the fields on a larger scale, so that some particles 
contribute to the density and momentum field of every grid point.
However, in doing so information is lost on the velocity field at 
all scales smaller than the smoothing scale.
Unlike the case of the density field, where we could subtract the 
effect of the smoothing scheme by de-convolving $P_\delta(k)$ in 
Fourier space, here we deal with a ratio of convolved fields,
which does not correspond to a simple multiplication in Fourier
space. Hence the deconvolution from the interpolating scheme is not
possible for the velocity field. This sets a limit on the resolution 
of the velocity power spectrum at small scales.

In our case, to define the velocity fields of our simulations we
evaluated the density and momentum density fields using TSC
interpolation onto a 5 Mpc grid, followed by Gaussian smoothing
of each with a kernel of size $7$ Mpc before the ratio of fields is
taken.  Since in linear theory only the longitudinal component 
of the velocity field $\vec v_\parallel$ (defined by the irrotationality 
condition: $\vec \nabla \times \vec v_\parallel =0$) is related to the
density fluctuation field $\delta(\vec x)$, we considered only the 
power spectrum of $\vec v_\parallel$. Fig. 10 compares 
$P_v(k)$ for P3M144, P3M256 and MAP8x144, superimposed on the linear 
prediction. The difference in amplitude between the latter and the power
spectra of the three simulations at high wavenumbers is an effect of the
filtering of the velocity field of the former.
\placefigure{fig10}
We can see from the Figure how $P_v(k)$ for the \MAP simulation shows
some amplitude difference over the spectra of both P3M144 and P3M256 on
scales smaller than about $25$ Mpc. This may again be related to the
sampling problem discussed before, or to the invalidity of the linear
approximation, but our current understanding of the effects of smoothing 
on the velocity field is too limited to allow a definite interpretation.

 From the simulations we also evaluated the pairwise velocity dispersion 
$\sigma_{v,12}$ as a function of pair separation. This is defined as the 
second central moment of the velocity field: given pairs of particles
with velocities $\vec v_1$ and $\vec v_2$, separated by a distance 
$r_{12}$, the parallel component of the pairwise velocity dispersion is
\begin{equation}
\sigma_{v,12 \parallel} = \left\langle\left[(\vec v_2 -\vec v_1)\cdot 
\hat r_{12}\right]^2\right\rangle^{1/2}.
\end{equation}
where $\hat r_{12} \equiv \vec r_{12}/|\vec r_{12}|$. 
\placefigure{fig11}
Since the value of $\sigma^2_{v,12 \parallel}(r)$ is determined by
the power associated with density fluctuations on scales $\lambda 
\mincir 1/r$, the pairwise velocity dispersion is a suitable 
statistic to estimate the small scale velocity power of a simulation.
Fig. 11 plots $\sigma_{v,12 \parallel}$ for our three simulations;
the three curves agree with each other to better than 10\%.
 From this figure our conclusion is that the \MAP has not changed the
velocity field significantly on small scales. Small differences
between the curves are found also for P3M256 and may just reflect 
statistical fluctuations. The global agreement of this statistic 
contrasts somewhat with the different amplitudes of the velocity 
power spectrum between P3M144 and MAP8x144.
This difference is not fully explained, but it could again be due
to nonlinear effects in the \MAP: the original simulation (L = 100 Mpc)
might still be too small to guarantee a sufficient linear evolution
for its fundamental modes, with $\lambda = L$. Mode coupling would then
propagate to small scales any change in the large scale power.
Put another way, the pairwise velocity seems more robust than the
velocity power spectrum in measuring the small scale power.

\section{Summary and Conclusions}

We have proposed a new method to add to an $N$-body simulation
the large scale power associated with scales larger than the volume
in which the simulation is performed. We made use of the Zel'dovich
approximation (Zel'dovich, 1970) to change each particle's position
and velocity according to the extra power introduced.  We tested the
method using a simulation of standard cold dark matter, which we
called P3M144. It had been evolved in a cube of $100$ Mpc on a side 
by means of a P$^3$M code within $144^3$ collisionless particles.
We replicated the simulation to fill a cube of side $800$ Mpc and added to 
it the power associated with fluctuations up to scales $\lambda=800$ Mpc.
We compared this enlarged simulation, named MAP8x144, with the
original P3M144 simulation and with a larger simulation in a
volume of $640$ Mpc on a side called P3M256.

We showed both visually and by means of several statistics how the density 
and velocity field are modified by the addition of long waves: velocities are 
increased and structures are created with characteristic scales larger 
than the original box size. The rms velocity of a particle  is
$v_{rms}=768$ km s$^{-1}$ in P3M144, $v_{rms}=827$ km s$^{-1}$ 
for MAP8x144, and $v_{rms}=804$ km s$^{-1}$ for P3M256. 
The equivalent figures for the density field show that the \MAP slightly
enhances the preexisting clustering.

Our analysis of several statistics shows the effects of long waves
in nonlinear simulations.  The \MAP procedure assumes that long and
short waves evolve independently and that the former are describable by
the Zel'dovich approximation.
However, our results suggest that there may be some transfer of power 
between long and short wavelengths in our simulations, for example
in the power spectra. If so, for accurate results our method may require
stronger conditions than those met in our simulations.
While a detailed study of this problem is beyond the scope of this
paper, we can try to shed more light 
on this point by examining Fig. 2b, which refers to the
simulation P3M256, performed on a cube of $640$ Mpc on a side.

We can identify three intervals of wavenumbers corresponding to 
different behaviors of the growth rates.
The first interval corresponds to fluctuations with 
$\log (k$~Mpc$) \geq -0.8$ (i.e. $\lambda \leq 40$ Mpc); these 
grow faster than linear in all plotted outputs, and their growth 
becomes faster at later times, defining what is usually called the
nonlinear regime.
A second intermediate interval is approximately 
$-1.5 \leq \log (k$~Mpc$) \leq -0.8$ 
(corresponding to scales between $200$ Mpc and $40$ Mpc); fluctuations
in this range grow slightly more slowly than the linear theory 
prediction, the effect becoming most visible at the latest time $a=0.7$. 
Finally, fluctuations with wavenumber $\log (k$~Mpc$) \leq -1.5$ 
($\lambda \magcir 200$ Mpc) maintain a strict linear growth (to within
1\%) at all times.
The analogous plot for P3M144 (shown in Fig. 2a) may suggest a similar
behavior at least for a=0.7, but unfortunately the growth rates are
much more unstable, due to the smaller size of the simulation,
so that a definite interpretation is not possible.

The existence of these three regimes suggests that some coupling
exists between the modes in the intermediate interval of wavenumbers 
and the modes in the nonlinear regime, with a transfer of power from 
the former to the latter. Transfer of power between long and short 
wavelength modes is consistent with the results of Jain and 
Bertschinger~(1994): using second order perturbative calculations
for a CDM-like spectrum, they found that mode coupling cause a slight
suppression of $P(k)$ at small $k$, and a significant enhancement at
high $k$ compared to the linear prediction, with the transition region
occurring where the spectral slope is $n\mincir-1$ 
(that is $k \magcir 0.1$ Mpc) respectively.
If this interpretation is correct, then the size of P3M144 ($100$ Mpc 
on a side) is slightly too small to perform the mode substitution,
because the longest waves in the simulation are still weakly coupled
with shorter wavelengths.
Therefore by subtracting the longest waves from P3M144 using our
displacement field technique we have also subtracted some power in the
density field from small, nonlinear scales. The small scale
power however is not given back with the addition of long waves up 
to $L_{big}$. In fact, the new large scale power was not present during 
the simulation, but is added randomly at the end of time evolution,
and so has no chance to dynamically enhance or suppress small-scale
waves.
Fortunately, even if some mode coupling affects the present example, 
it does not represent a limit of the method but just of the simulation
we used to apply the \MAP, so the conclusions we drew on the method
are still valid.
One obvious way to verify this hypothesis is to run the \MAP starting 
with a simulation originally performed on a larger volume, for example
$200$ Mpc on a side. In this case we would expect to see no
significant power transfer.

On a completely different issue, we would like to stress the 
point that our end product is not equivalent to a real simulation 
evolved from initial conditions on a comparable scale. In fact, 
outside the substituted region of Fourier space the MAP8x144 
simulation samples high wavenumbers exactly like P3M144. That is, 
the density of Fourier modes there is not as high as in an
$N$-body simulation actually performed on $800$ Mpc on a side. 
Moreover, some small scale periodicity may still be present in the
final result, even if modulated by the large scale waves. 

The \MAP can also be applied to any catalog of dark matter halos, or to
any class of objects that can be defined in a simulation. In that case, 
however, one cannot trace back the Lagrangian position of the objects, 
since they contain different particles, and are defined during -- or 
after -- time evolution.
What one does instead is to apply in Equations (\ref{erem}) and (\ref{eadd})
the displacement field obtained by the original particle distribution which 
is parent to the halo catalog. That is, the displacement field of Equations 
(\ref{ldisp}) and (\ref{edisp}) is obtained from the particles.

Finally, the application of the \MAP described in this paper concerns the 
study of very large scale density and velocity fields. However, other 
applications are possible, with focus on different aspects. In fact, the 
procedure that we have described here does not require at all the use of 
a box as big as the longest added wavelengths. 
Once the mode removing step is performed,
one can interpolate the added displacement field to an arbitrary volume
of the simulation. For example, in a study that does not require using a
very large simulated volume, one can introduce the large-scale power in 
the original simulation without taking any replica. In such a case all the
particles of the original simulation may be used, to preserve the initial
high definition and resolution.

\acknowledgments
We would like to thank Jim Frederic for providing the P3M256 simulation,
and Sabino Matarrese and Simon White for useful comments and suggestions 
on an earlier version of the manuscript. We wish to thank the referee,
Michael Strauss, for a helpful referee's report. Thanks are also given to
the National Center for Super-computing Applications which provided our
computer time.  Financial support was provided by NSF grant AST-9318185
and for G.T. also by the Italian MURST and by an EC-HCM fellowship.


\centerline{\bf Figure Captions}

\figcaption{
Left: pictorial representation of the Fourier space for a 
periodic simulation on a cubic box with side $L$. The plot shows one 
plane of a three-dimensional distribution.
Each small solid square corresponds to a position where the power spectrum 
$P_\delta(k)$ is sampled by the initial conditions given to the simulation. 
The dashed grid divides the space in cubes of equal volume $(\Delta k)^3$ 
where $k=(2 \pi/L)$, and each cube is associated with a mean power per mode 
$P_\delta(k)(\Delta k)^3$.
Right: pictorial representation of the Fourier space of a \MAP simulation:
adding power from wavelengths larger than the original $L$ corresponds
to placing a finer sampling of the power spectrum around $\vec k=\vec 0$.
See text for further discussion.\label{fig1}} 

\figcaption{
Rate of growth for the density fluctuation power spectra of P3M144 (Fig. 2a)
and P3M256 (Fig. 2b). Growth is plotted as a function of $k$ and is
normalized to the growth predicted by linear theory. Deviations from the 
linear approximation show as departures of the curve from a unity
value.\label{fig2}} 

\figcaption{
\MAP applied to a Lagrangian slice of particles originally on a grid.
We placed $200 \times 200$ test particles on a regular two dimensional lattice
of side $800$ Mpc, and displaced them by mean of the (three dimensional) long 
wave displacement field $\psi_+(\vec x)$ described above in the text.
The field $\psi_+(\vec x)$ has been generated using the same Fourier modes 
that will be added to P3M144, i.e. using only the power coming from scales
between  $800$ Mpc and roughly $50$ Mpc. The left panel shows a projection of
the  entire slice; the right panel is a blow up of $200$ Mpc on a side taken
from the left panel.\label{fig3}}

\figcaption{
In a box of $800$ Mpc on a side particles were laid in order 
to mimic a two dimensional network of one dimensional structures. The
particles were initially distributed in the $x-y$ plane along $8 + 8$
filaments stretching across the box in both the $x$ and $y$
directions, as indicated by the dashed lines.
The filaments have uniform density along their axis; each contains 5000 
particles. The broken solid line shows how this network of particles
has been modulated by the action of the long waves depicted in
Fig. 3. The cut refers to a slice of thickness 20 Mpc. The sections of
network apparently missing are just displaced along the line of sight
($z$ axis), enough to make them fell out of the slice here shown.\label{fig4}}

\figcaption{
Displacement exerted by the \MAP (in Mpc) on one of the
filaments of Fig. 4.
Left Panel: three components of the displacement field versus the coordinate 
along the filament. Central Panel: amplitude of the total displacement along 
the filament. Right Panel: blow up of the steepest part of Central
Panel.\label{fig5}}

\figcaption{
Eulerian slice taken from P3M144 (Fig. 6a), MAP8x144 (Fig. 6b) and P3M256
(Fig. 6c). Each slice is $640\times 640\times 10$ Mpc$^3$. Fig. 6a only 
replicates subsamples of particles from P3M144; Fig. 6b shows the same
Eulerian slice shown in Fig. 6a, after long waves have been added up to 
a scale of $800$ Mpc. Fig. 6c is a comparable slice from the true 
$N$-body simulation P3M256, performed in a volume of $640$ Mpc on a
side. Note that, since in Fig. 6c only one particle every two from
P3M256 is plotted, a pattern appears as if particles were aligned 
along chains. This is just due to the grid used in the initial 
conditions.
\label{fig6}}

\figcaption{
Density fluctuation distribution functions for P3M144 (solid line), MAP8x144
(dotted-dashed) and P3M256 (long dashed). Differential (a) and cumulative (b)
distributions are plotted. In each case the density has been calculated by 
interpolating the particle positions on a regular cubic lattice with intergrid 
spacing of $5$ Mpc. A TSC interpolating scheme was used.
See the text for discussion.\label{fig7}} 

\figcaption{
Density fluctuation power spectra for the three simulations. The plotted
curves have been corrected for shot noise and have had the mass assignment
scheme deconvolved.\label{fig8}} 

\figcaption{
Differential and cumulative distributions for the magnitude of the
peculiar velocities in P3M144, MAP8x144 and P3M256. Note the increasing of
velocities with the box size. Velocities in MAP8x144 are even larger than 
those in P3M256, suggesting that nonnegligible contributions come from
scales beyond $640$ Mpc even for a standard CDM cosmological model.
\label{fig9}} 

\figcaption{
Power spectra for the longitudinal component of the velocity field, for the 
three simulations. A Gaussian smoothing filter of size 7 Mpc was applied 
to every field to define the velocity field also in underdense regions.
The slight amplitude difference between MAP8x144 and P3M144 at high frequencies
is not significative, as the velocity power spectrum is not well defined on 
small scales. Unlike $P_\delta(k)$, here no shot noise subtraction or window 
deconvolution has been applied.\label{fig10}} 

\figcaption{
Parallel component of the pairwise velocity dispersion of the particles
in the simulations.\label{fig11}} 

\end{document}